\documentclass[acmsmall]{acmart}

\usepackage{acronym} 
\usepackage{algorithm} 
\usepackage{algorithmic}
\usepackage{booktabs}
\usepackage{clrscode3e}
\usepackage{hyperref}
\usepackage{multirow}
\usepackage{graphicx}
\usepackage{diagbox}
\usepackage{subfigure}
\usepackage{textcomp}
\usepackage{xcolor}
\usepackage{tabularx}
\usepackage{threeparttable}
\usepackage{adjustbox}
\usepackage{tabularx}

\acrodef{MLP}{Multi-Layer Perception}
\acrodef{HR}{Hit Ratio}
\acrodef{NDCG}{Normalized Discounted Cumulative Gain}
\acrodef{MCR}{Many-to-one Cross-domain Recommendation}
\acrodef{CR}{Cross-domain Recommendation}
\acrodef{CSR}{Cross-domain Sequential Recommendation}
\acrodef{MCRPL}{Many-to-one Cross-domain Recommendation via Prompt Learning}
\acrodef{NMCR}{Non-overlapping Many-to-one Cross-domain Recommendation}

\AtBeginDocument{%
  \providecommand\BibTeX{{%
    \normalfont B\kern-0.5em{\scshape i\kern-0.25em b}\kern-0.8em\TeX}}}

\setcopyright{acmcopyright}
\copyrightyear{2023}
\acmYear{2023}
\acmDOI{XXXXXXX.XXXXXXX}

\acmJournal{TOIS}
\acmVolume{00}
\acmNumber{0}
\acmArticle{000}
\acmMonth{0}

\begin{document}

\title{MCRPL: A Pretrain, Prompt \& Fine-tune Paradigm for Non-overlapping Many-to-one Cross-domain \\ Recommendation}



\author{Hao Liu}
\email{liuhaosdnu@gmail.com}
\affiliation{
  \institution{Shandong Normal University}
  \city{Jinan}
  \state{Shandong}
  \country{China}
  \postcode{250358}
}

\author{Lei Guo}
\authornotemark[1]
\email{leiguo.cs@gmail.com}
\affiliation{
  \institution{Shandong Normal University}
  \city{Jinan}
  \state{Shandong}
  \country{China}
  \postcode{250358}
}

\author{Lei Zhu}
\authornote{Corresponding Author.}
\email{leizhu0608@gmail.com}
\affiliation{
  \institution{Shandong Normal University}
  \city{Jinan}
  \state{Shandong}
  \country{China}
  \postcode{250358}
}

\author{Yongqiang Jiang}
\email{yongqiang@robot.soc.i.kyoto-u.ac.jp}
\affiliation{
  \institution{Kyoto University}
  \city{Kyoto}
  \country{Japan}
}

\author{Min Gao}
\email{gaomin@cqu.edu.cn}
\affiliation{
  \institution{Chongqing University}
  \city{Chongqing}
  \country{China}
}

\author{Hongzhi Yin}
\email{h.yin1@uq.edu.au}
\affiliation{
  \institution{The University of Queensland}
  \city{Brisbane}
  \country{Australia}
}

\renewcommand{\shortauthors}{Liu, and Guo et al.}

\begin{abstract}
Cross-domain Recommendation (CR) is the task that tends to improve the recommendations in the sparse target domain by leveraging the information from other rich domains. Existing methods of cross-domain recommendation mainly focus on overlapping scenarios by assuming users are totally or partially overlapped, which are taken as bridges to connect different domains. However, this assumption does not always hold since it is illegal to leak users' identity information to other domains. Conducting Non-overlapping MCR (NMCR) is challenging since 1) The absence of overlapping information prevents us from directly aligning different domains, and this situation may get worse in the MCR scenario. 2) The distribution between source and target domains makes it difficult for us to learn common information across domains. To overcome the above challenges, we focus on NMCR, and devise MCRPL as our solution. To address Challenge 1, we first learn shared domain-agnostic and domain-dependent prompts, and pre-train them in the pre-training stage. To address Challenge 2, we further update the domain-dependent prompts with other parameters kept fixed to transfer the domain knowledge to the target domain. We conduct experiments on five real-world domains, and the results show the advance of our MCRPL method compared with several recent SOTA baselines. 
Moreover, Our source codes have been publicly released\footnote{https://github.com/Estadio-c/MCRPL}.
\end{abstract}



\maketitle

\section{Introduction}
Due to the success of \acf{CR}~\cite{buchong1,buchong2,buchong3,buchong4,buchong5,buchong6,buchong7,buchong8} in alleviating the data sparsity issue by leveraging the information from other rich domains, \ac{CR} has been extensively investigated by recent studies. Moreover, as users' interaction behaviors are often organized as sequences, the \ac{CSR} task has become an emerging research topic.
According to the number of domains involved, existing \ac{CR}s can be divided into dual-target~\cite{dual_1,dual_2,dual_3}, multi-target~\cite{multi_1,multi_2}, and single-target~\cite{singe_1,singe_2,singe_3,singe_4} \ac{CR} tasks. 
For single-target task, Pan et al.~\cite{singe_4} exploit users' binary auxiliary preference data to reduce the impact of data sparsity issues in rating predation tasks. Zhang et al.~\cite{many_to_one1} extract group-level knowledge from multiple source domains and apply domain adaptation techniques to eliminate domain shift, aligning users and projects to maintain consistency of knowledge.  For dual-target task, Sun et al.~\cite{psj-net} use a gating mechanism to filter out valuable auxiliary information from mixed user behaviors to enhance sequential representation and then migrates it to another domain. For multi-target task, Cui et al.~\cite{multi_1} alleviate sparsity by aggregating neighbors from users or items in multiple domains through graph construction and efficiently transfer information across domains. 
Among them, methods on the single-target \ac{CR} task are the typical techniques that tend to alleviate the data sparsity issue in the target domain by leveraging the information in one or multiple source domains.
In this work, we will take the multiple source-based sing-target \ac{CR} (i.e., \ac{MCR}) are our learning objective, since it is more practical in many real-world systems. For example, it is a common practice to involve more domains in the \ac{CR} scenario, as users usually have diverse interests.
Moreover, as \ac{MCR} needs to transfer information from multiple source domains to the target, it is more challenging than only considering a single source domain.

\begin{figure*}
    \centering    \includegraphics[width=13cm]{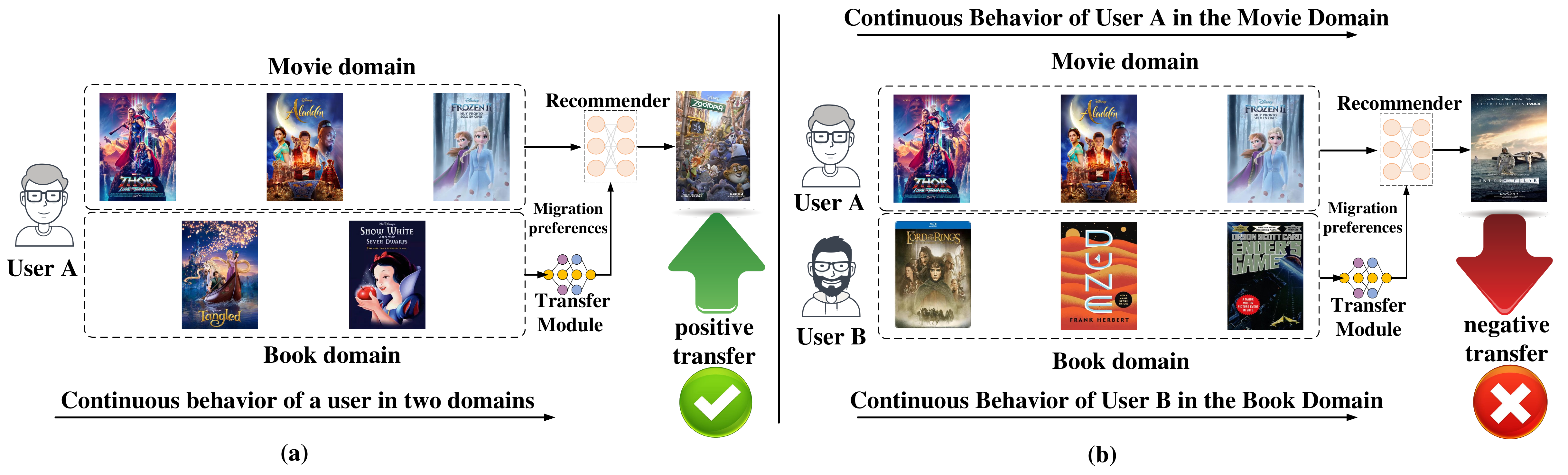}
    \caption{An example that illustrates the drawbacks of traditional cross-domain methods in the non-overlapping scenarios.}
    \label{fig:introduction}
\end{figure*}

However, pioneer works on \ac{MCR} are mainly based on the overlapping assumption that users in different domains are totally or partially overlapped. With this assumption, they can connect different domains by viewing these users as bridges, and further conduct domain migration from source to the target domain.
But this assumption does not always hold, because user identities are not allowed to be leaked to other platforms for privacy reasons. It is unrealistic to assume that we can simultaneously acquire users' behaviors in all platforms, and the non-overlapping condition brings new challenges to the \ac{CR} tasks.
An illustrative example is shown in Fig.~\ref{fig:introduction},
where user A is present in both the Movie and Book domains, and we can access her interactions at the same time (in Fig.~\ref{fig:introduction} (a)).
Assuming user A has watched three movies in the movie domain, i.e., Thor, Aladdin Lamp, and Frozen.
As all these movies are about science fiction and animation, we can recommend movies within these genres to her.
Moreover, as we also know user A has recently read Le Pei and Snow White in the Book domain, we can further update our recommendations by recommending Zootopia to her -- one of the most popular animated movies at that time.
However, the \ac{CR} results may worsen when the overlapping assumption does not hold.
As shown in Fig.~\ref{fig:introduction} (b), users A and B are non-overlapped across the Movie and Book domains. It is infeasible to directly align their preferences across domains since A and B have entirely different interest preferences.

Although previous works~\cite{non_overlaping_1,non_overlaping_2,non_overlaping_3,non_overlaping_4} have proposed relevant solutions to the non-overlapped issue,
their methods are only applicable to the dual-target or single-target \ac{CR} tasks, and only few of them investigate \ac{MCR} under the non-overlapping scenario.
We argue that conducting non-overlapping \ac{MCR} is a challenging task, due to: 1) 
The absence of overlapping information prevents us from directly aligning users' interests in different domains, and this situation may get worse when there are multiple source domains, since domains may have completely different data distributions.
It is challenging to migrate information across domains without explicit domain alignment.
2) There is a significant difference in data sparsity between the source and target domains, making traditional \ac{CR} methods prone to generate negative transferals, because of the inability to learn domain information from sparse data.
Despite there have been studies on improving the recommendations in the sparse target domain by utilizing dense domain data, they are all based on overlapping assumptions.
The non-overlapping \ac{MCR} task is largely unexplored.

To overcome the above challenges, we focus on the \ac{NMCR} task and propose a prompting paradigm for \ac{MCR}, namely \ac{MCRPL}, which mainly consists of a prompt layer and sequence modeling layer.
To model users' cross-domain interests under the non-overlapping scenario, we devise a prompt layer with shared prompts to avoid direct domain alignment. We then pre-train it by utilizing the data from both source and target domains.
By doing this, we can integrate the common knowledge between domains into prompts, which will be further fine-tuned to meet the data distribution of the target domain.
Moreover, as we pre-train the prompts and sequence encoder by all the source and target domains, the data sparsity issue can be largely alleviated by transferring the domain knowledge via the shared prompts for each item.
In this work, the prompts in \ac{MCRPL} are composed of domain-shared and domain-specific prompts, where the former tends to model the common knowledge between different domains and will be frozen during the fine-tuning stage, and the latter aims at learning the information specific to the target domain in the fine-tuning stage.
Furthermore, to enable prompts to learn different aspects of the domain knowledge, an orthogonal loss function is applied between them.

Our main contributions can be summarised as follows:
\begin{itemize}
    \item We propose a prompt learning paradigm with a two-stage training strategy for \ac{NMCR}, and jointly consider the non-overlapping and data sparsity characteristics in the many-to-one recommendation task.
    \item We devise a prompt layer with shared prompts to avoid direct domain alignment, and then pre-train them with all the domains to face the data sparsity challenge. We fine-tune the pre-trained model by freezing the sequence encoder and domain-agnostic prompts for knowledge transfer.
    \item We develop two types of prompts, i.e., the domain-agnostic and domain-specific prompts, to learn different aspects of domain knowledge with an orthogonal loss function.    
    \item We conduct extensive experiments on two real-world datasets, HVIDEO and MIXED, to demonstrate the superiority of our proposal.
\end{itemize}
\section{Related Work}
In this section, we consider three types of recommendation methods, i.e., sequential recommendation, cross-domain recommendation, and prompt learning-based recommendation, as our related works. We then discuss the differences between our approach and other related \ac{CR} methods.

\subsection{Sequential Recommendation}
The purpose of sequential recommendation~\cite{seq_1,seq_2,seq_3,seq_4} is to use continuous user behavior records to recommend products that are most likely to interact in the future. Due to the lack of product features~\cite{yin1,yin2,yin3}, it has been widely used in various commercial platforms. The earliest sequence recommendation method was based on ~\cite{markov}, which calculated the transition probability between states and used this model to predict and recommend the next action and item based on the current state. With the increasing application of Recurrent neural networks, Hidasi et al.~\cite{GRU4Rec} use GRU units to model user sequences, surpassing the method based on the Markov chain in performance and practicality. Subsequent studies such as Qiu et al.~\cite{graph_3} and Zhu et al.~\cite{graph_1} improve the predictive ability of the model for future interacting items by leveraging the powerful modeling capability of graph neural networks. Kang et al.~\cite{SASRec} first introduce the attention mechanism into sequence recommendation, and compared with previous methods, the model shows faster computational speed and better performance. After the widespread adoption of attention-based models, subsequent research efforts have made numerous optimizations to enhance their performance. Zhou et al.~\cite{S3-Rec} utilize existing data correlation to obtain self-supervised signals, and enhance data representation through pre-training methods to alleviate data sparsity. Xie et al.~\cite{CL4SRec} propose multiple data augmentation methods and combined them with comparative learning techniques to obtain higher-quality user representations. 
One of the major problems of the traditional sequential recommendation methods is that they only consider the single domain, and cannot directly migrate user preferences to other domains to assist cross-domain recommendation, which is the main purpose of this work.

\subsection{Cross-domain Recommendation}
The cross-domain recommendation aims to enhance the recommendation performance of the target domain by utilizing information from other domains. According to the ultimate goal of the task, cross-domain recommendation methods can be roughly divided into three categories: single-~\cite{singe_1,singe_2,singe_3,singe_4}, dual-~\cite{dual_1,dual_2,dual_3,pai-net,reinforcement,guo2022time}, and multi-target~\cite{multi_1,multi_2}. For single-target tasks, Liu et al.~\cite{singe_2} propose a stein path alignment method that utilizes the rating and auxiliary representation of the source domain to improve the recommendation performance of the target domain. Pan et al.\cite{singe_3} discover the coordinates of users and items in the auxiliary data matrix and migrated them to the target domain through coordinate system offset to alleviate the impact of data sparsity. The purpose of dual-target and multi-target recommendations is to simultaneously improve the performance of two or more domains. For dual-target tasks, Liu et al.~\cite{dual_3} propose a graph collaborative filtering network, which uses public users as a bridge to realize the two-way transfer of cross-domain knowledge and improve the performance of the two domains. For multi-target task, Cui et al.~\cite{multi_1} introduce a shared graph structure to model information from multiple domains, greatly simplifying the process of cross-domain modeling and enhancing the flow of domain information. The above method can largely address the problem of cross-domain recommendation, but most of them are based on overlapping entities, which is difficult to satisfy in the real world. Subsequently, a small amount of work~\cite{multi_2,Non-overlapped,Partially-Overlapped} has studied cross-domain methods under non-overlapping or partially overlapping entity conditions. Krishnan et al.~\cite{multi_2} share domain invariant components to enable the model to learn higher-quality user and item representations in the sparse domain, achieving one-to-many transmission under non-overlapping conditions. Liu et al.~\cite{Non-overlapped} propose a collaborative filtering algorithm based on attribute alignment under the condition of no overlapping users, reducing the domain discrepancy between users and items. 

Currently, there are only a few studies have investigated the \ac{MCR} task. For example, Moreno et al.~\cite{many_to_one2} propose a transfer learning technique to learn the relationships between multiple source and target domains in non-overlapping scenarios and transfer knowledge from multiple dense domains to the sparse domain. Zhang et al.~\cite{many_to_one1} conduct domain adaptation through flexible constraints in matrix factorization to eliminate domain bias and improve recommendation performance in the sparse domain without relying on overlapping entities.
Though the above methods consider the non-overlapping scenario, they mainly focus on exploring users' rating preferences, and their sequential preferences are largely unexplored.

\subsection{Prompt Learning-based recommendation}
The prompt fine-tuning technique~\cite{Survey} is used to adapt large-scale pre-trained models to downstream tasks. At present, prompt design is mainly divided into soft prompts~\cite{soft-1,soft-2} and hard prompts~\cite{hard-1,hard-2}. For example, Schick et al.~\cite{hard-1} use of hard prompts made up of real words makes the model suitable for special scenarios. Due to the high cost of manually designing prompts, many subsequent work focuses on fine-tuning soft prompts, which are composed of continuous learnable embeddings. Zhou et al.~\cite{coop} use the clip model for downstream image recognition by optimizing only a small number of parameters for soft prompts. Currently, some work has applied prompt learning technology to recommendation systems~\cite{P5,personalized,explainable,fairness,rethinking,yin4}. Geng et al.~\cite{P5} use the T5~\cite{t5} model as the backbone network and unified various recommendation tasks in a shared framework by designing a large number of prompt words. Wu et al.~\cite{personalized} use user profiles to generate personalized soft prompts and combined them with comparative learning strategies to solve the cold-start problem. At present, the application of prompt learning in recommendation systems is still in its early stages. We believe that excellent prompt design can transfer public information from multiple domains to the target domain without using overlapping entities.

\subsection{Differences}
Our work has significant differences from previous studies: 1) Our \ac{MCRPL} is different from existing studies on \ac{NMCR}~\cite{many_to_one1,many_to_one2}, which align domains by inferring users' group-level preferences under the traditional training paradigm. But they fail to handle the distribution difference between source and target domains, resulting in sub-optimal results.
Moreover, all previous methods explore \ac{NMCR} on the basis of users' rating behaviors, and none of them consider this scenario on users' sequential preference, which is more practical in real-world systems.
In this work, we address \ac{NMCR} on users' sequential behaviours, and propose a prompting-based method as our solution. Specifically, to address the non-overlapping challenge, we resort to the pretrain, prompt \& fine-tune paradigm, and unify the domain distributions by pre-training the recommendation task universally.
2) Our method is different from existing prompt-based cross-domain recommendation methods. For example, Guo et al.~\cite{guo_prompt} devise a prompt learning-based cross-domain recommendation method, but their work is focused on the \ac{CR} task with only one source domain, and the multi-source \ac{CR} task is largely unexplored.
3) Our prompt-based training schema is also different from other typical prompting methods. Compared with manual-based prompt learning methods~\cite{P5}, our \ac{MCRPL} 
paradigm explores the automated prompt engineering method to fit the target domain better.
Compared with learnable prompts~\cite{personalized,plate}, our \ac{MCRPL} paradigm adaptively weights and aggregates different levels of contexts words through an elaborate-designed attention mechanism.

\section{METHODOLOGY}
In this section, we first present the overall framework of our prompt learning paradigm for \ac{NMCR}, and then detail how we implement our proposed solution \ac{MCRPL}. Table~\ref{tab:symbol} summarizes the main symbols and notations used in this work.
\begin{figure}
    \centering    \includegraphics[width=14.5cm]{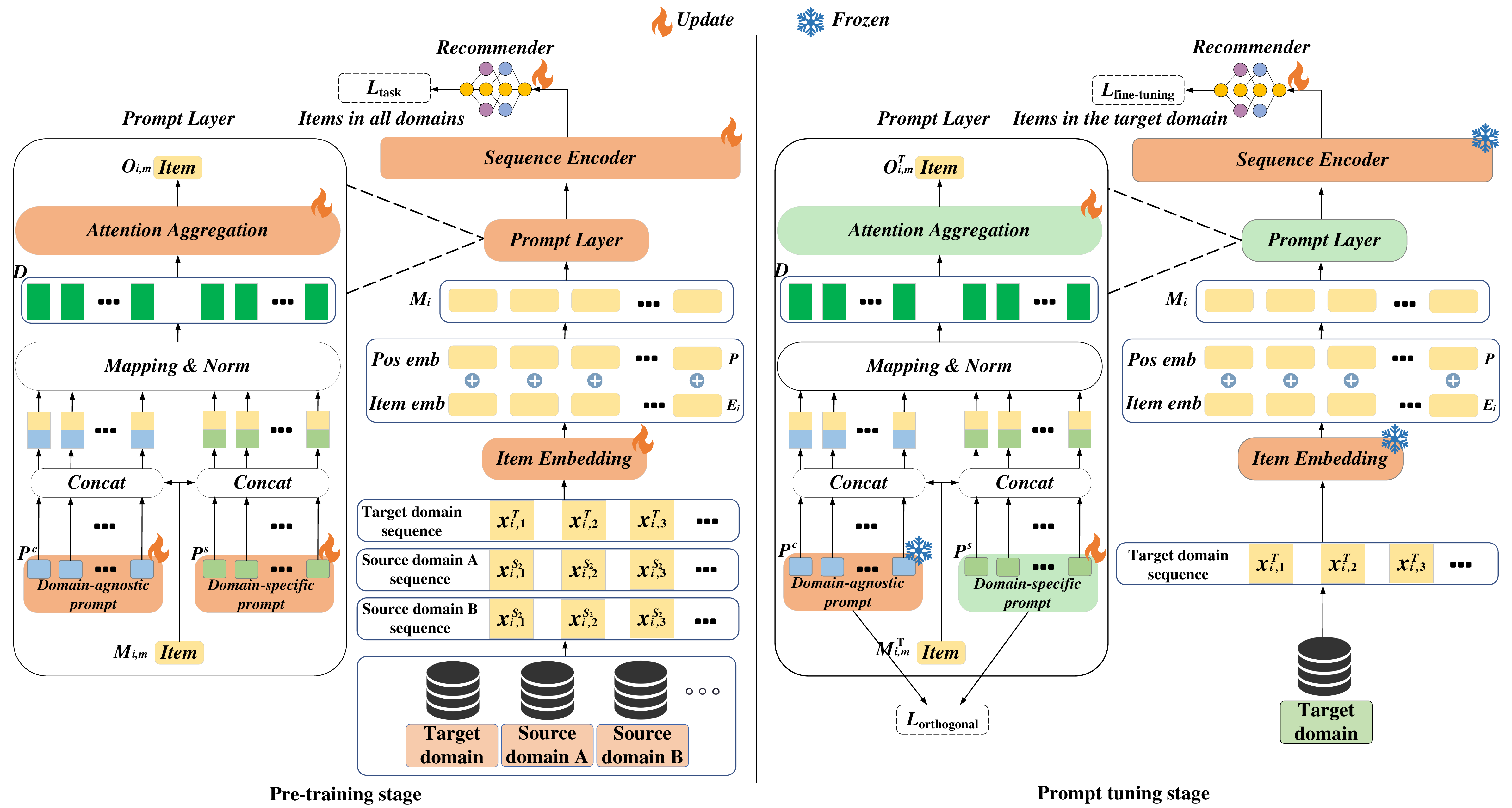}
    \caption{The overall system architecture of \ac{MCRPL}. For each item in the inputs, we enhance its representation by the prompt layer.}
    \label{fig:overview}
\end{figure}
\subsection{Preliminaries}
Suppose we have a sparse target domain $T$ and multiple dense source domains set $S = \{S_1, S_2, \ldots, S_d\}$, and no overlapping users and items exist among these domains.
Let $\mathcal{I}^T= \{x^T_1, x^T_2, ..., x^T_{n}\}$ be the item set in domain $T$, and $\mathcal{I}^{S_i}= \{x^{S_i}_1, x^{S_i}_2, ..., x^{S_i}_{n}\}$ be the item set in source domain $S_i$, where $\mathcal{I}^T \cap \mathcal{I}^{S_i} = \phi$.
We can access users' interaction sequences in source and target domains $\mathcal{X}_i^T=\{x_{i,1}^T, x_{i,2}^T, ..., x_{i,m}^T, ..., x_{i,M}^T,  y_i^T\}$ (take target domain as an example), where $x_{i,m}^T$ denotes the m-th item interacted item by user $u_i$,  $M$ is the length of the sequence, and $y_i^T$ is the target of the sequence.
Note that, as users are non-overlapped among domains, we can only get their behaviors in a single domain.
Moreover, as users' interactions are often organized into sequences, we tend to make sequential recommendations, that is, recommending the next item that will be interacted with by the target user.
Furthermore, as we assume the target domain is spare, we can only get limited user behaviors, which means we cannot accurately model users' interests based solely on their behaviors in the target domain.
The target of \ac{NMCR} is to improve the sequential recommendation in the sparse target domain by leveraging the information in dense source domains under the non-overlapping condition.
More formally, the task of \ac{NMCR} can be defined as follows: 

\textbf{Input:} The behavior sequence $\{x_{i,1}^F, x_{i,2}^F, ..., x_{i,m}^F, ..., x_{i,M}^F\}, F \in \{S,T\}$ in both source and target domains. 

\textbf{Output:}  The probability $P\left(y_i^T\mid \mathcal{X}_i^T\right)$ of recommending $y_i^T$ as the next item to be consumed in the target domain T.
\begin{table}[htbp]
  \centering
  \caption{Summary of the main symbols and notations used in this work.}
  \begin{tabular}{ll}
    \toprule
    \textbf{Symbol} & \textbf{Notation} \\
    \midrule
    $T$ & Target domain \\
    $S$ & Multiple dense source domains set $\{S_1, S_2, \ldots, S_d\}$ \\
    $|\textbf{A}|$ & The number of training sequences in both the source and target domains \\
    $|\textbf{T}|$ & The number of training sequences in the target domain \\
    $L_p$ & Number of prompts \\
    $d$ & The embedding dimension of prompts and items \\
    $\mathcal{I}^T$ & Item set $\{x^T_1, x^T_2, ..., x^T_{n}\}$ in target domain $T$ \\
    $\mathcal{I}^{S_i}$ & Item set $\{x^{S_i}_1, x^{S_i}_2, ..., x^{S_i}_{n}\}$ in source domain $S_i$ \\
    $x_{i,m}^T$ & The m-th item of user $u_i$ interaction \\
    $\mathcal{X}_i^T$ & Continuous behavior sequence $\{x_{i,1}^T, x_{i,2}^T, ..., x_{i,m}^T, ..., x_{i,M}^T, y_i^T\}$ of target domain $u_i$ \\
    $\boldsymbol{E}^{T}_{i}$ & Embedding Matrix of Continuous Behavior of $u_i$ \\
    $\boldsymbol{P}$ & Position embedding matrix \\
    $\boldsymbol{P}^{C}$ & Domain-agnostic prompts composed of $L_p$ contextual words embedding \\
    $\boldsymbol{P}^{S}$ & Domain-specific prompts composed of $L_p$ contextual words embedding \\
    \bottomrule
  \end{tabular}
  \label{tab:symbol}
\end{table}
\subsection{Overview}
\textbf{Motivation.}
In the non-overlapping scenario, it is not feasible to leverage traditional domain adaption methods for cross-domain recommendation, since users are totally disjoint, and there is no common users can be take as bridges to connect different domains. The inability to accurately align different domains is the main reason for negative transfer in traditional methods.
And, this situation may get worse in the \ac{MCR} scenario, because of the differences in data distribution between different domains.

To conduct information transfer without domain alignment, we tend to first learn the global common view among domains, and then take them as prior knowledge to enhance the recommendations in the sparse target domain.
Different from priors methods, we resort to the prompt learning paradigm with a two-phrase training method. 
Specifically, we devise two kinds of prompts, which are learned by pre-training the recommendation model on the data from all the source and target domains. This allows us to integrate the common information into prompts and take them as the prior knowledge of the target domain.
To transfer the domain information from multiple source domains to the target domain, we further fine-tuning the pre-trained model by freezing the domain-shared prompts and sequence encoder, and adapt them to the target domain.

As \ac{MCRPL} does not need direct domain alignment, it avoids the possible irrelevant information as users often have diverse interests, and many of them are unrelated to the target domain.
Moreover, since in our fine-tuning stage, the domain-agnostic prompt is fazed and only the domain-specific prompt is fine-tuned, we can leverage the common knowledge to improve the recommendations in the target domain while preserving its specific characteristics.

\textbf{Overall Framework.}
The system architecture of \ac{MCRPL} is shown in Fig.~\ref{fig:overview}, which mainly consists of a prompt layer and a sequence encoder along with a two-stage training schema. 
1) To learn transferable domain information without any overlapping information, we tend to encode the common domain knowledge through the shared prompts in the prompt layer, which are responsible for storing the users' shared interests in different domains.
Moreover, to model different aspects of the domain information, we design the prompts as domain-agnostic and domain-specific prompts, where the former targets learning the knowledge related to all the domains and the latter aims at modeling the information relevant to specific domains (an orthogonal loss is also applied).
All these prompts are integrated into each item representation, and aggregated by an attention network (the details can be seen in Section \ref{Prompt_layer}).
2) The sequence encoder tends to capture users' preferences from their sequential behaviors -- a transformer-style modeling method, which takes the output of the prompt layer as input, and outputs the corresponding representation of the current sequence (the details can be seen in Section \ref{Sequence_Encoder}).
3) Our training process is separated into a pre-training stage and a fine-tuning stage. In the pre-training stage, we pre-learn the prompts and encoders via all the domain data to model the common knowledge among domains under the non-overlapping constraint.
In the fine-tuning stage, we freeze the parameters in domain-agnostic prompts and sequence encoder and fine-tune the domain-specific prompts on the target domain to transfer the domain knowledge from multiple source domains, while keeping the characteristics of the target domain (the details can be seen in Section \ref{Two-stage}).

\subsection{Data Initialization}
To alleviate the data sparsity in the target domain and unify the feature spaces of item representations in different domains, we first mix all the source and target domain data, and leverage them to pre-train the basic recommendation model by predicting the next items that users are inclined to interact in all domains.

In our setting, all the users' interactions are organized into fixed-length sequences with $l_t$ as their maximum length.
For sequences shorter than $l_t$, we apply the padding policy on the left side of them to enable the input to have a uniform length. For sequences longer than $l_t$, we only retain the last $l_t$ behaviors in them as the training samples.
We represent items in all the domains by an item embedding matrix $\boldsymbol{F^T}\in \mathbb{R}^{n \times d}$, where $n$ denotes the number of items and $d$ is the dimension of the item embedding.
We represent the initial vectors of a input sequence $\mathcal{X}_i^T=\{x_{i,1}^T, x_{i,2}^T, ..., x_{i,m}^T, ..., x_{i,M}^T\}$
by a lookup operation on $\boldsymbol{F^T}$, and denote it by $\boldsymbol{E}^{T}_{i}=\operatorname{emb}\left(x_{i,1}^T, x_{i,2}^T, ..., x_{i,m}^T, ..., x_{i,M}^T\right)$ (take the target domain T as an example).

To encode the position of each item in their corresponding sequences, we represent each position by a learnable position embedding matrix $\boldsymbol{P^T}\in \mathbb{R}^{l_t \times d}$, and the position embedding of the m-th position item is represented as  $\boldsymbol{P}^T_m = \operatorname{{PositionEmbedding}}(pos_m)$ ($pos_m$ is the positional index of the m-th item).
Thus, the initial embedding of the input sequence can be enhanced by their positions, which are denoted as:
\begin{equation}        
            \boldsymbol{M}^T_i=\left[\begin{array}{c}       
            \boldsymbol{E}^T_{i,{1}}+\boldsymbol{P}^T_{1} \\ 
            \boldsymbol{E}^T_{i,{2}}+\boldsymbol{P}^T_{2} \\
            \ldots \\      
            \boldsymbol{E}^T_{i,{m}}+\boldsymbol{P}^T_{m} \\
            \ldots \\      
            \boldsymbol{E}^T_{i,{l_t}}+\boldsymbol{P}^T_{l_t}
            \end{array}\right]
            \label{init}
\end{equation}


\subsection{Prompt-enhanced Item Representation \label{Prompt_layer}}
To learn the common knowledge from disjoint domains, we further enhance the items by the public prompts, which are expected to embed the shared information among domains.
Moreover, to model different aspects of the domain information, we devise domain-agnostic and domain-specific prompts in the prompt layer.
By introducing prompts, we can not only model the shared domain knowledge but also transfer them from pre-trained source domains to the target domain to enhance the recommendations within it. 
Our prompt design is shown in Fig.~\ref{fig:overview}, which consists of $L_p$ domain-agnostic and $L_p$ domain-specific prompt context words, where $m$ and $n$ are hyper-parameters and will be fine-tuned in experiments.
First of all, each item $x_{i,m}^T$ within the given sequence $\mathcal{X}_i^T$ will be separately added to all the context words. Then, the enhanced items are then aggregated by an attention mechanism, since context words may have different weights in representing the domain information.

\subsubsection{Domain-agnostic Prompts} 
The domain-agnostic context words of prompts are expected to learn domain-irrelevant features that are shared by all the source and target domains.
As in \ac{NMCR}, both users and items are non-overlapping across domains, we tend to first map the common domain knowledge into a shared feature space, and treat them as the prior knowledge of the target domain during the fine-tuning stage. By this, we can transfer the domain knowledge without domain alignment.

The design of domain-agnostic prompts is shown in Fig.~\ref{fig:overview}.
Suppose there are $L_p$ context words in $\boldsymbol{P}^{C} \in \mathbb{R}^{L_p \times d}$, and each of them $\boldsymbol{P}^{C}_n$ with dimension $d$.
In our setting, $\boldsymbol{P}^{C}$ is shared by all the items in both source and target domains.
We use it to embed common knowledge across domains, which can also be considered as a clustering representation.
Note that, the domain-agnostic prompt is only optimized during the pre-training stage, and we will freeze the parameters within it to keep the common characteristics in the fine-tuning stage.

\subsubsection{Domain-specific Prompts}
To learn domain-related features, we devise another {\color{red}{$L_p$}} context words to represent the domain-specific prompts $\boldsymbol{P}^{S} \in \mathbb{R}^{L_p \times d}$.
Similar to domain-agnostic prompts $\boldsymbol{P}^C$, domain-specific prompts are also shared by the items in both source and target domains.
But different from $\boldsymbol{P}^C$, $\boldsymbol{P}^S$ is optimized in both stages. That is, as one of the main purposes of the pre-training stage in MCRPL is to embed the common domain knowledge into prompts (i.e., both domain-agnostic and domain-specific prompts tend learn the same aspects of domains), which are then be treated as prior knowledge to enhance domain adaption in the fine-tune stage.
However, since the source domain and the target are quite different, we also need to model domain-agnostic and domain-specific features. Hence, we further add the orthogonal loss in the fine-tuning stage to let the pre-learned knowledge can be adapted to the target domain.
\begin{align}  
&\mathcal{L}_{orthogonal}=\left\|\boldsymbol{P}_{1}^{C^{\top}} \boldsymbol{P}_{1}^{S}\right\|_{2}+\left\|\boldsymbol{P}_{2}^{C^{\top}} \boldsymbol{P}_{2}^{S}\right\|_2+\cdots+ \left\|\boldsymbol{P}_{L_p}^{C^{\top}} \boldsymbol{P}_{L_p}^{S}\right\|_2,
    \label{diff}
\end{align}
where $\boldsymbol{P}^C$ is frozen during the fine-tuning stage. $||\cdot||_{2}$ is the Euclidean norm of a vector.
The orthogonal loss between $\boldsymbol{P}^C$ and $\boldsymbol{P}^S$ tend to make them orthogonal. In this way, $\boldsymbol{P}^S$ can model the domain features that are only related to the target domain T, enabling the pre-trained model can be well adapted to T.

\subsubsection{Prompt Aggregation}
Both types of prompts are further concatenated with every item within the given sequence to empower items to have the ability to model the cross-domain information (as shown in Fig.~\ref{fig:overview}).
\begin{equation}
\boldsymbol{D}_{n}=\operatorname{Norm}\left[\left(\boldsymbol{M}^B_{i,m} \| \boldsymbol{P}_n^{B}\right) \boldsymbol{W_{1}}\right], n=1,2,...,2*L_p, B \in C,S \\
\label{concat}
\end{equation}
where $\|$ is a vector concatenation operation, $\boldsymbol{W}_1 \in \mathbb{R}^{2d\times d}$ are trainable parameters, and we use it to reduce the dimensionality of $\boldsymbol{M}^T_{i,m} \| \boldsymbol{P}_n^{T}$ to $d$. Norm ($\cdot$) is the row-normalized function, $2* L_p$ is the total number of the context words in prompts.

Then, we aggregate the resulting prompts $D_n$ by an attention mechanism, since different contexts words should have different contributions to the corresponding item.
Specifically, we choose $\boldsymbol{D}_{2*L_p}$ as the aggregation base point, calculate weight coefficients, and perform aggregation, which is represented by the formula:
\begin{align}
    & \boldsymbol{\alpha}_{{n,k}}=\frac{(\boldsymbol{D}_n \boldsymbol{W}^{Q}) (\boldsymbol{D}_k \boldsymbol{W}^{K})^{\top}}{\sqrt{d_k}}, \label{attention_alpha}\\
    & \boldsymbol{\beta}_{n,k}=\frac{\exp (\alpha_{n,k})}{\sum_{j=1}^{2*L_p} \exp (\alpha_{n,j})}, \\
    & \boldsymbol{z}_n=\sum_{j=1}^{2*L_p}\beta_{n,j}D_j,
    \label{attention}
\end{align}
where $\boldsymbol{W}^Q \in \mathbb{R}^{d\times d}$, $\boldsymbol{W}^K \in \mathbb{R}^{d\times d}$ are two linear projections matrices. $\boldsymbol{\alpha}_{n,k}$ is the dot-product attention between $\boldsymbol{D}_n$ and $\boldsymbol{D}_k$, and $\boldsymbol{\beta}_{n,k}$ is its normalized version. Then, the attentively aggregated representation $\boldsymbol{z}_n$ can be denoted by the weighted sum over the values $\boldsymbol{D}$.
\begin{align}
    & \boldsymbol{O}_{i,m}=\operatorname{Pool}\left(\left[\boldsymbol{z}_{1} ; \boldsymbol{z}_{2} \ldots ; \boldsymbol{z}_{2*L_p}\right]\right)
    \label{pool}
\end{align}
\begin{algorithm}[b]
\footnotesize
\caption{$\proc{The training process of \ac{MCRPL}}.$}
\begin{algorithmic}[1]
\label{alg:training_process}
\REQUIRE  User sequences $\mathcal{X}_i^S=\{x_{i,1}^S, x_{i,2}^S, ..., x_{i,m}^S, ..., x_{i,M}^S,y_i^s\}$ from the set of the source domain $S = \{S_1, S_2, \ldots, S_d\}$, User sequences $\mathcal{X}_i^T=\{x_{i,1}^T, x_{i,2}^T, ..., x_{i,m}^T, ..., x_{i,M}^T,y_i^T\}$ from the domain T, 
the number of epochs, batch size, learning rate, and the hyper-parameters of \ac{MCRPL};
\ENSURE The prediction probability for the candidate items in the target domain;
\STATE Initialize the model parameters;
\STATE Shuffle the input sequences (i.e., $S$ and $T$) in both domains; 
\STATE \textbf{Pre-training phase};
\FOR{each epoch}
  \FOR{($\mathcal{X}_i$) in \{${S,T}\}$}
  \item Initialize the embedding of sequence $\mathcal{X}_i$ using Eq.~(\ref{init});
  \item Obtain the enhanced sequence embedding$\boldsymbol{O}_i$ through the aggregation layer using Eqs.~(\ref{concat}), (\ref{attention}) and (\ref{pool});
   \item Obtain the representation of the sequence using the sequence encoder according to Eqs.~(\ref{seq_1}) and (\ref{seq_2}).
   \item Calculate the recommendation probabilities for all items in all domains according to Eq.~(\ref{predict1}).
   \item Calculate the loss and optimize the parameters of the entire model according to Eq.~(\ref{task1}).
   \ENDFOR
\ENDFOR
\STATE \textbf{Fine-tuning phase};
\STATE Freeze domain-agnostic prompt and sequence encoder;
\FOR{each epoch}
  \FOR{($\mathcal{X}^T_i$) in \{${T}\}$}
  \item Initialize the embedding of sequence $\mathcal{X}^T_i$ using Eq.~(\ref{init});
  \item Obtain the enhanced sequence embedding $\boldsymbol{O}^T_i$ through the aggregation layer using Eqs.~(\ref{concat}), (\ref{attention}) and (\ref{pool});
   \item Obtain the representation of the sequence using the sequence encoder according to Eqs.~(\ref{seq_1}) and (\ref{seq_2}).
   \item Calculate the domain orthogonal loss according to Eq.~(\ref{diff}).
   \item Calculate the recommendation probabilities for all items in domain $T$ according to Eq.~(\ref{predict2}).
   \item Calculate the loss and optimize the parameters of the entire model according to Eq.~(\ref{eq:training_loss}).
   \ENDFOR
\ENDFOR
\end{algorithmic}
\end{algorithm}
\subsection{Sequence Encoder \label{Sequence_Encoder}}
In this work, we leverage the encoder blocks in the Transformer~\cite{attention} as the sequence encoder for its advantage and effectiveness in sequence modeling.
To learn users' sequential preferences on their interacted items, we first transform the enhanced items $\boldsymbol{O}_{i}=\left\{\boldsymbol{o}_{i, 1}, \boldsymbol{o}_{i, 2}, \cdots, \boldsymbol{o}_{i, m},\cdots,\boldsymbol{o}_{i,l_t}\right\}$ into queries $Q$, keys $K$, and values $V$ ( $Q= O W^Q$, $K= O W^K$, $V= O W^V$), where $\mathbf{W}^{Q}, \mathbf{W}^{K}, \mathbf{W}^{V} \in \mathbb{R}^{d\times d}$ are the projection matrices.
Then, we can get the high-dimensional transformation of the input by an attention layer:
\begin{align}
&\operatorname{Attention}(\mathbf{Q}, \mathbf{K}, \mathbf{V})=\operatorname{softmax}\left(\frac{\mathbf{Q K}^{T}}{\sqrt{d}}\right) \mathbf{V}\\
&\boldsymbol{U}_{i}=\operatorname{Attention}\left(\boldsymbol{O}_{i} \mathbf{W}^{Q}, \boldsymbol{O}_{i} \mathbf{W}^{K}, \boldsymbol{O}_{i} \mathbf{W}^{V}\right),
\label{seq_1}
\end{align}
where $\mathbf{Q K}^{T}/\sqrt{d}$ is the dot-product attention between queries and keys, and $\boldsymbol{U}_i$ is the output of the attention layer, denoting the result of the attentive aggregation over the values $\boldsymbol{V}$.
Then, to endow the extraction with nonlinearity, we further apply a point-wise two-layer feed-forward neural network to $\boldsymbol{U}_i$:
\begin{align}    &\boldsymbol{seq}_{i,l_t}=\operatorname{ReLU}\left(\boldsymbol{U}_{i} \boldsymbol{W}_2+\boldsymbol{b}_1\right) \boldsymbol{W}_3+\boldsymbol{b}_2,
\label{seq_2}
\end{align}
where $\boldsymbol{W}_2$, $\boldsymbol{W}_3$ is the weight matrix of a feed-forward network, $\boldsymbol{b}_1$, $\boldsymbol{b}_2$ is the bias, and ReLU is the activation functions.
The last enhanced item of each sequence aggregates the temporal information of the entire sequence through a multi-layer attention network, which best reflects the user's recent interests. Therefore, we leverage the last item representation $\boldsymbol{seq}_{i,l_t}$ as the whole sequence embedding.
 
\textbf{Pre-training Objective.} 
We define $\textbf{A} = \{T, S_1, S_2, \ldots, S_d\}$ as the set of all source and target domains. To enable the learned parameters in different domains within a unified feature space, we train them with a single learning objective, that is, we pre-train the recommender to make predictions by selecting the items in all domains. The probability of predicting the next item is defined as:
 \begin{equation}
      P\left(y_{i+1} \mid \mathcal{X}_i^{\textbf{A}}\right)=\operatorname{softmax}\left(\boldsymbol{W}_{all} \cdot \boldsymbol{seq}_{i,l_t}+\boldsymbol{b_3}\right)
      \label{predict1}
 \end{equation}
where $\boldsymbol{seq}_{i,l_t}$ is the sequence embedding, $\boldsymbol{b}_3$ is the bias term, and $\boldsymbol{W}_{all}$ denotes the item embeddings in both source and target domains.

Then, we can achieve our loss function by calculating the following cross-entropy:
 \begin{equation}
     \mathcal{L}_{task}=-\frac{1}{|\textbf{A}|} \sum_{\mathcal{X}^{\textbf{A}}_i \in \textbf{A}} \sum_{x_{i,m}^{\textbf{A}} \in \mathcal{X}^{\textbf{A}}_i} \log P\left(y^\textbf{A}_{i+1} \mid \mathcal{X}^{\textbf{A}}_i\right),
     \label{task1}
 \end{equation}
where |\textbf{A}| represents the number of sequences in all domains, Eq.~(\ref{task1}) is the objective function for the first stage of our model training, which can be optimized by the Adam algorithm.
 
\subsection{The Prompt Tuning Stage \label{Two-stage}}
To transfer domain knowledge across 
disjoint domains, we first pre-train our recommender by the data in all the source and target domains and embed the common knowledge into the shared prompt contexts, which can also be viewed as the prior knowledge of the target domain. 
Then, to deliver the domain information from the source to the target domain, we further fine-tune the pre-trained model and adapt it to meet the distribution of the target domain by the prompting technique.
The fine-tuning process on the target domain can be described as the following formula:
\begin{equation}
       P\left(y_{i+1} \mid \mathcal{X}_i^{\textbf{T}}\right)=\operatorname{softmax}\left(\boldsymbol{W}_{T} \cdot \boldsymbol{seq}_{i,l_t}+\boldsymbol{b_3}\right) 
      \label{predict2}
\end{equation}
\begin{equation}
     \mathcal{L}_{fine-tuning}=-\frac{1}{|\textbf{T}|} \sum_{\mathcal{X}^{\textbf{T}}_i \in \textbf{T}} \sum_{x_{i,m}^{\textbf{T}} \in \mathcal{X}^{\textbf{T}}_i} \log P\left(y^\textbf{T}_{i+1} \mid \mathcal{X}^{\textbf{T}}_i\right),
     \label{tesk2}
 \end{equation}
Concretely, in the fine-tuning stage, we fix the parameters in domain-agnostic prompts, item embeddings, and sequence encoder, and adapt the pre-trained model by only fine-tuning the domain-specific prompts.
The context words in the domain-specific prompts are expected to learn the domain-related information, while the fixed domain-agnostic parameters are aiming at enhancing this process by providing the related prior knowledge.
By doing this, the prompts can capture commonalities between domains while preserving
the individuality of the target domain. 
To encourage domain-specific prompts to better capture the differences between the source and target domains, we add a orthogonal loss $\mathcal{L}_{orthogonal}$ to these two kinds of prompts.
The definition of $\mathcal{L}_{orthogonal}$ is shown in Eq. (\ref{diff}).

Then, we can achieve the learning objective of the fine-tuning stage:
\begin{align}    \mathcal{L}=\mathcal{L}_{fine-tuning}+\lambda\mathcal{L}_{orthogonal }
    \label{eq:training_loss}
\end{align}
The freezing of domain-agnostic prompts and embedding layer parameters ensures the preservation of domain public information. By adding domain orthogonal loss and fine-tuning domain-specific prompts, the personalized preferences of target domain users can also be preserved. In the fine-tuning stage, we optimize domain private prompts using the Adam algorithm according to Eq.~(\ref{eq:training_loss}).
The training process of our final objective function is shown in Algorithm \ref{alg:training_process}.

\subsection{Prediction Process}
In the prediction stage, all the operations are conducted in the fined-tune stage of \ac{MCRPL}.
Specifically, to make predictions for the users in the target domain, we first feed their sequences to the data initial layer to obtain items' initial representations $\boldsymbol{M}^T_i$.
Then, we can get the enhanced item embeddings $\boldsymbol{O}_i$ by leveraging the prompt layer, where the domain-agnostic and domain-specific prompts are integrated with each of the items within the current sequence.
After that, we learn users' sequential interests $\boldsymbol{seq}_{i,l_t}$ by the pre-trained sequence encoder 
and finally get the predictions for every candidate item (as shown in Eq. (\ref{predict1})).
\section{Experimental Setup}
In this section, we first introduce the research questions and then present the datasets and evaluation protocols utilized in our experiments. Finally, we detail the baselines and how we implement our method.

\subsection{Research Questions}
We fully evaluate our proposal by answering the following research questions:
\begin{itemize}
    \item[\textbf{RQ1}] How does the proposed \ac{MCRPL} method perform compared with the state-of-the-art recommendation baselines?
    \item[\textbf{RQ2}] How do the key components of \ac{MCRPL}, i.e., domain-agnostic prompts, domain-specific prompts, the two-stage training strategy, and domain orthogonal loss contribute to its recommendation performance?
    \item[\textbf{RQ3}] What are the effects of the fine-tuning strategies on the performance of \ac{MCRPL}?
    \item[\textbf{RQ4}] How does \ac{MCRPL} perform with different hyper-parameters?
    \item[\textbf{RQ5}]  How is the training efficiency and the scalability of \ac{MCRPL} when processing large-scale data?
\end{itemize}

\subsection{Datasets}

We conduct experiments on a MIXED-domain (VIDEO, MOVIELENS, WIKIPEDIA)~\cite{SASRec} and HVIDEO~\cite{pai-net} to evaluate our proposed method. 
VIDEO, MOVIELENS, and WIKIPEDIA are the datasets that are separately from three different platforms (VIDEO is the target domain).
Specifically, VIDEO is a video domain in the AMAZON dataset\footnote{https://jmcauley.ucsd.edu/data/amazon/}, which includes user ID, item ID, timestamp, rating, item description, item image, and other information. This dataset has sparsity and significant data offset, making it very suitable for the target domain and requiring significant robustness of the model.
The MOVIELENS dataset\footnote{https://grouplens.org/datasets/movielens/} is a widely used movie rating dataset collected and published by the GroupLens Research laboratory. This dataset contains 1 million rating records from all users for 6040 movies. The average user behavior sequence of this dataset is particularly dense, making it suitable for the source domain.
WIKIPEDIA\footnote{https://github.com/pmixer/SASRec.pytorch/tree/master/data/} is built based on the user browsing behavior and page content information of Wikipedia websites. This dataset contains a large amount of user browsing behavior data, including information such as user clicks on different pages, browsing time, and browsing order. The average sequence length of this dataset is also dense, making it suitable for use as the source domain.
By selecting domains from three different platforms, we can assure users and items are totally disjoint to satisfy the non-overlapping characteristic of the \ac{NMCR} task.
To meet the sequential characteristic of \ac{NMCR}, the users' interactions in all domains are organized into sequences in chronological order. 
Since the average user sequence length of MOVIELENS and VIKIPEDIA is relatively dense, we will not modify the original data of these two datasets.
For the VIDEO dataset, to meet the data sparsity scenario, we randomly select 5 to 8 items at the end of each user sequence. Our target is to transfer information from the relatively dense source domains (MOVIELENS and VIKIPEDIA) to improve the recommendations in the sparse target domain (VIDEO). 

To evaluate our performance on two disjoint domains, experiments on HVIDEO are further conducted~\cite{pai-net}. 
HVIDEO is a smart TV data released by Ma et al.~\cite{pai-net} that records users' TV watching histories on the E domain and V domain. 
Concretely, the V domain includes users' video-watching behaviors toward TV dramas, movies, animations, etc. E-domain records users' watching interactions on educational videos from primary school to high school.
As users are overlapped in this dataset, we conduct the random shuffle operation on these two domains to break the connections of them to satisfy the non-overlapping characteristic of \ac{NMCR}.
To meet the sparse requirement, we also randomly select 5 to 8 items at the end of each user sequence on the target domain (E-domain).
Our target is to improve the recommendations in the E-domain by leveraging the information in V-domain.
The statistics of our datasets are shown in Table 1.

\begin{table}
    \centering
    \caption{The statistics of our HVIDEO and MIXED datasets.}
    \begin{tabular}{lcccccc}
    \toprule
    & \multicolumn{3}{c}{HVIDEO}&\multicolumn{3}{c}{MIXED} \\
    \cmidrule{1-7}
    \multirow{1}[1]{*}{\#domain}& \multicolumn{3}{c}{E (Sparse)} & \multicolumn{3}{c}{VIDEO (Sparse)}\\
    \#Items &\multicolumn{3}{c}{3,389} &\multicolumn{3}{c}{23,715}\\
    \#Interactions &\multicolumn{3}{c}{629,209} &\multicolumn{3}{c}{155,240}\\
    \#Avg. sequence length & \multicolumn{3}{c}{5.45} & \multicolumn{3}{c}{5.02} \\
    \#sequence number & \multicolumn{3}{c}{115,385} & \multicolumn{3}{c}{30,901} \\
    \cmidrule{1-7}
    \multirow{1}[1]{*}{\#domain}& \multicolumn{3}{c}{V (Dense)} & \multicolumn{3}{c}{MOVIELENS (Dense)}\\
    \#Items &\multicolumn{3}{c}{16,431} &\multicolumn{3}{c}{3,416}\\
    \#Interactions &\multicolumn{3}{c}{1,970,378} &\multicolumn{3}{c}{999,611}\\
    \#Avg. sequence length & \multicolumn{3}{c}{14.66}  & \multicolumn{3}{c}{165.50} \\
    \#sequence number & \multicolumn{3}{c}{115,385} & \multicolumn{3}{c}{6,040} \\
    \cmidrule{1-7}
    \multirow{1}[1]{*}{\#domain}& \multicolumn{3}{c}{---} & \multicolumn{3}{c}{WIKIPEDIA (Dense)}\\
    \#Items & \multicolumn{3}{c}{---} & \multicolumn{3}{c}{1,000} \\
    \#Interactions  & \multicolumn{3}{c}{---} & \multicolumn{3}{c}{151,990} \\
    \#Avg. sequence length & \multicolumn{3}{c}{---} & \multicolumn{3}{c}{38,40} \\
    \#sequence number & \multicolumn{3}{c}{---} & \multicolumn{3}{c}{3,958} \\
    \bottomrule
    \end{tabular}
    \label{tab:dataset_statistics}
\end{table}

\subsection{Evaluation Protocols}
For the MIXED dataset, as users do not have too much historical interactions, we do not further split these interactions into smaller sequences. Instead, we represent each user by all her interactions.
Then, we split these sequences into training, validation, and testing by random sampling at ratios of 75\%, 15\%, 10\%. For the HVIDEO dataset, we first conduct the random shuffle operation on all the users with the chronological order of each user's interactions unchanged. Then, we split each sequence into smaller segments based on the chronological order of her interactions.
To ensure that users' future behavior is not used to predict their current behaviors, we directly take 75\%, 15\% and 10\% of all sentences in order according to the time order of their division as the training, validation and testing sets, respectively.

All the remaining behaviors within the current sequence are treated as the inputs (denoted as $\mathcal{X}_i^F, F \in \{T,S\}$).
Our sequential recommendation task is to predict the next item ($y_i^T$) from the candidates based on $\mathcal{I}^T$.
As selecting parts of items (e.g., 100) as the candidates as the evaluation strategy are not unstable~\cite{sampled}, 
we deem all the items in our dataset as the candidate set and tend to make recommendations from all of them.
Since users usually make decisions based on the top-ranked items, we chose two widely ranking-based metrics MRR@$K$ and Reacll@$K$ to evaluate our method, where the values of $K$ are set as 3, 5, 10, 20 on MRR@$K$ and 3, 5, 10 on Reacll@$K$.
\begin{itemize}
 \item Recall@$K$: This metric measures the fraction of the desired items that can be successfully covered amongst the top-$K$ recommender items. 
\item MRR@$K$: As Recall does consider the actual rank of the items, we further introduce MRR (Mean Reciprocal Rank) to evaluate the top-$K$ recommendations. 
   MRR is the average of reciprocal ranks of the relevant items, of which the reciprocal rank is set to zero if the ground truth item is not in the top-$K$ recommendation list.
   Compared with Recall, MRR takes into account the rank of the items, which is vital in scenarios where the item ranking orders matter.
\end{itemize}

\subsection{Baselines} \label{baseline}
We compare with two types of methods, i.e., sing-domain methods and cross-domain methods, to evaluate the performance of \ac{MCRPL}.

Single-domain approaches:
\begin{itemize}
 \item \textbf{Pop}: This is a simple yet competitive baseline, which provides a top-$K$ recommendation list based on the popularity of items in the training set, and always recommends the same item to each user.
 \item \textbf{BPR-MF}~\cite{bpr}: This is a classic matrix decomposition algorithm that decomposes users and items into potential spaces to calculate similarity for the recommendation. Since this method cannot handle sequence information, we average pooling all items that appear in the sequence to represent the sequence and apply it to sequence recommendation. 
 \item \textbf{GRU4Rec}~\cite{GRU4Rec}: This method exploits GRU as the sequence encoder to capture users' continuous behaviors. This is the first method that introduces recurrent neural networks into sequential recommendations. In our experiments, we use a single-layer GRU network as the backbone model.
 \item \textbf{SASRec}~\cite{SASRec}: This is the first method that utilizes the masked multi-head attention mechanism to model sequences. It is a competitive baseline, due to its effectiveness in modeling users' sequential behaviors.
\item \textbf{S3Rec}~\cite{S3-Rec}: 
This is a sequential recommendation method that leverages the inherent data correlations to obtain self-supervised signals for model pre-training. But as this method needs additional attribute information, we compare our method to a variant of this method by removing the losses depending on items' attributes during the pre-training phase.
\item \textbf{FMLP-Rec}~\cite{Filter}: 
This method devises a learnable filter for sequence recommendation tasks via borrowing the idea of filtering algorithms in signal processing.
\item \textbf{ICLRec}~\cite{ICLRec}: 
This method aims to enhance the performance and robustness of the sequential recommendation model using contrastive self-supervised learning by exploring the user's latent intents.
\end{itemize}
As the following methods are proposed for overlapped users in two domains, we disturb the orders of the domain data to remove the connections between domains for fair comparisons.

Cross-domain approaches:
\begin{itemize}
\item \textbf{\textbf{$\pi$}-net}~\cite{pai-net}: 
This method conducts the sequential recommendations by developing a parallel GRU network, where a cross-domain transmission unit is devised to enhance the recommendations in both domains by shared users.
 \item \textbf{PSJ-net}~\cite{psj-net}: This is an improved method over \textbf{$\pi$}-net, which  
 further develops a parallel split connection network for learning role-specific representations and uses gating mechanisms to filter out user role information that may be useful for another domain from mixed user behaviors.
 \item \textbf{DAGCN}~\cite{dagcn}: 
 This is a graph-based solution for \ac{CR} tasks, where a domain-aware graph convolutional network with two novel message passing and aggregation strategies is proposed.
 \item \textbf{PLCR}~\cite{guo_prompt}: This method utilizes the automated prompt engineering paradigm to design prompt components that contain domain information at different levels, dynamically adapting to specific domains.
\end{itemize}

\subsection{Implementation Details}
We implement our \ac{MCRPL} method on the PyTorch platform and accelerate the
training process via a GeForce GTX TitanX GPU. We use the Xavier Normal initialization method~\cite{init}
to initialize all the parameters in \ac{MCRPL}. 
Our training process is divided into two stages: pre-training and the fine-tuning stages. In our training process, we load the pre-trained model parameters and fine-tune the target domain-related parameters to make \ac{MCRPL} learn from the pre-trained prior knowledge.
In both the pre-training and fine-tuning stages, we use the Adam algorithm as the optimizer (the learning rate is set as 0.001 in the pre-training stage).
We further utilize a learning rate scheduling strategy 
to find the optimal parameters in the fine-tuning phase.
We set the maximum learning rate as 0.001 and the warm-up step as 10 epochs.
For both source and target domains, the batch size is set as 128. 
For the length of the inputs, we set the maximum sequence length as 6 for the spare target domain and 30 for the dense source domain.
For the sequence encoder in \ac{MCRPL}, we utilize a single-layer transformer architecture with only one head. The embedding dimensions for both items and prompts are set as 100.
For the hyper-parameter $\lambda$ in the objective function (i.e., Eq. (\ref{eq:training_loss})), we explore its value in the range of [0.01, 0.07] with step 0.01.

For the hyper-parameters in our compared baselines, we set their values as follows: 1) For single domain methods (i.e., BPR-MF, GRU4Rec, SASRec, S3Rec, FMLP-Rec, ICLRec), we only use the target domain for training and testing. 
For BPR-MF, since it cannot directly handle sequential data, we average pool all the embeddings in the sequence to represent the sequence, and set the learning rate to 0.001, and batch size to 128.
For GRU4Rec,  we use a single-layer GRU network as the backbone model. For SASRec, 
We set the learning rate to 0.001 and the number of layers and heads in the backbone network to 2 and 1 respectively, following the original paper. For S3Rec, We set the number of layers and heads in the attention network to 2, and the batch size for the pre-training and fine-tuning stages to 200 and 256 respectively, following the original paper. For FMLP-Rec, We set the hidden size to 64, the learning rate to 0.001, and the batch size to 256. For ICLRec, we set the number of layers and heads in the backbone network both as 2, following their original paper
2) For cross-domain methods (i.e., $\pi$-net, PSJ-net, DAGCN, PLCR), their training requires paired user behavior sequences from different domains. To simultaneously satisfy the conditions of non-overlapping and paired user sequences, we adopt the method of randomly extracting user interactions from other domains to form paired sequences to train cross-domain baselines. For $\pi$-net, we set the dimension of the embedding layer and the hidden size of the GRU to 90, and the batch size to 64. For PSJ-net, we keep the same parameter settings as $\pi$-net. For DAGCN, we set the hidden size to 100 and the batch size to 128. For PLCR, we refer to its original paper and set the learning rate and batch size as 0.0001 and 128, respectively.

\begin{table*}
  \centering
  \caption{Experimental Results (\%) on HVIDEO and MIXED Datasets. The best results are highlighted with bold text and the second-best results are highlighted with an underline.}
  \begin{threeparttable}
    \begin{tabular}{cccccccc}
      \toprule
      \multirow{4}{*}{\textbf{Methods}} & \multicolumn{7}{c}{ \textbf{HVIDEO} (E domain $\leftarrow$ V domain)} \\
      \cmidrule{2-8}
      & \multicolumn{3}{c}{Recall} & \multicolumn{4}{c}{MRR} \\
      \cmidrule{2-8}
      & @3 & @5 & @10 & @3 & @5 & @10 & @20 \\
      \midrule
      POP & 4.288 & 5.931 & 10.537 & 2.683 & 3.061 & 3.639 & 4.071 \\
      BPR-MF~\cite{bpr} & 4.962 & 6.106 & 8.105 & 2.793 & 3.218 & 4.576 & 4.792 \\
      GRU4Rec~\cite{GRU4Rec} & \underline{13.507} & 16.931 & 24.536 & 11.149 & 11.926 & 12.926 & 13.732 \\
      SASRec~\cite{SASRec} & 13.173 & 16.968 & 26.415 & 10.391 & 11.242 & 12.457 & 13.521 \\
      S3Rec~\cite{S3-Rec}  & 12.832 & 16.219 & 25.624 & 11.567 & 12.157 & 12.769 & 13.982 \\
      FMLP-Rec~\cite{Filter} & 10.560 & 15.294 & 25.869 & 7.042 & 8.107 & 9.479 & 10.644 \\
      ICLRec~\cite{ICLRec} & 11.244 & 15.849 & \underline{26.499} & 10.825 & 10.461 & 12.725 & 13.226 \\
      \midrule
      $\pi$-net~\cite{pai-net} & 13.188 & 16.703 & 25.847 & 10.569 & 11.368 & 12.563 & 13.580 \\
      PSJ-net~\cite{psj-net} & 12.278 & 15.988 & 25.659 & 10.126 & 10.982 & 12.168 & 13.676 \\
      DAGCN~\cite{dagcn} & 12.779 & 15.584 & 26.154 & 10.147 & 11.784 & 12.195 & 13.046 \\
      PLCR~\cite{guo_prompt} & 13.204 & \underline{17.492} & 26.329 & \underline{12.741} & \underline{13.149} & \underline{14.374} & \underline{15.435} \\
      \midrule
      \textbf{MCRPL} & \textbf{17.347*} & \textbf{20.233*} & \textbf{26.756*} & \textbf{15.224*} & \textbf{15.812*} & \textbf{16.605*} & \textbf{17.227*} \\
      \midrule
      \multirow{4}{*}{\textbf{Methods}} & \multicolumn{7}{c}{ \textbf{MIXED} (VIDEO domain $\leftarrow$ MOVIELENS domain, WIKIPEDIA domain)  } \\
      \cmidrule{2-8}
      & \multicolumn{3}{c}{Recall} & \multicolumn{4}{c}{MRR} \\
      \cmidrule{2-8}
      & @3 & @5 & @10 & @3 & @5 & @10 & @20 \\
      \midrule
      POP & 1.666 & 2.133 & 2.866 & 1.355 & 1.448 & 1.535 & 1.627 \\
      BPR-MF~\cite{bpr} & 1.832 & 2.189 & 2.567 & 1.204 & 1.328 & 1.498 & 1.564 \\
      GRU4Rec~\cite{GRU4Rec} & 3.099 & 4.199 & 7.031 & 1.822 & 2.224 & 2.589 & 2.812 \\
      SASRec~\cite{SASRec} & 3.251 & 4.532 & 7.665 & \underline{2.533} & 2.654 & 2.962 & 3.562 \\
      S3Rec~\cite{S3-Rec} & 3.232 & 4.499 & 7.763 & 2.260 & 2.574 & 2.821 & 3.305 \\
      FMLP-Rec~\cite{Filter} & 3.428 & 4.365 & 7.531 & 2.224 & 2.499 & 2.756 & 3.178 \\
      ICLRec~\cite{ICLRec} & 3.147 & 4.478 & 7.753 & 2.862 & \underline{3.145} & \underline{3.677} & \underline{3.962} \\
      \midrule
      $\pi$-net~\cite{pai-net}  & 2.685 & 3.985 & 6.731 & 1.556 & 1.988 & 2.175 & 2.489 \\
      PSJ-net~\cite{psj-net} & 2.432 & 4.257 & 6.894 & 1.432 & 1.758 & 2.006 & 2.256 \\
      DAGCN~\cite{dagcn} & 2.995 & 3.941 & 7.156 & 1.674 & 1.912 & 2.205 & 2.564 \\
      PLCR~\cite{guo_prompt} & \underline{3.824} & \underline{5.149} & \underline{8.257} & 2.497 & 2.877 & 3.286 & 3.751 \\
      \midrule
      \textbf{MCRPL} & \textbf{4.500*} & \textbf{6.465*} & \textbf{10.130*} & \textbf{3.327*} & \textbf{3.723*} & \textbf{4.199*} & \textbf{4.507*} \\
      \bottomrule
    \end{tabular}
    \begin{tablenotes}
      \footnotesize
      \item Significant improvements over the best baseline results are marked with * (t-test, $p$< .05).
    \end{tablenotes}
  \end{threeparttable}
  \label{tab:results}
\end{table*}

\section{Experimental Results (\textbf{RQ1})}
The comparison results on all the datasets are shown in Table~\ref{tab:results}. From these results, we can find that:
1) \ac{MCRPL} outperforms all baselines in all metrics, demonstrating the effectiveness of \ac{MCRPL} in addressing the data sparsity issue in the target domain.
2) The superior performance of \ac{MCRPL} over all single-domain methods indicates that the two-stage training strategy, combined with the prompt module, effectively transfers the common information of dense domains.
Although POP is simple, it can achieve a basic level of recommendation performance. The superior performance of MCRPL over POP indicates that modeling based solely on item popularity is not suitable for sparse domains. The higher performance of MCRPL over GRU4Rec, indicating the ineffectiveness of GRU in modeling the sparse sequences.
3) MCRPL shows significant improvements over attention-based single-domain methods (i.e., GRU4Rec, SASRec, S3Rec, FMLP-Rec, ICLRec), showing the advance of \ac{MCRPL} in enhancing the item embeddings and obtain higher-quality sequential representations. GRU4Rec outperforms attention-based methods on HVIDEO, indicating the significant impact of the data sparsity on such methods.
4) The performance gap between MCRPL and FMLP-Rec, demonstrating the capability of our two-stage training strategy in deducing the domain knowledge from source domains while preserving the individual preferences of the target domain. 
From this result, we also find that when the average sequence length of users is short, the performance of FMLP is lower than attention-based methods, indicating that filtering algorithms may not effectively address the issue of sparsity.
5) \ac{MCRPL} outperforms all cross-domain baselines (i.e., $\pi$-net, PSJ-net, DAGCN, PLCR), demonstrating the suitability of \ac{MCRPL} in transferring domain knowledge in the non-overlapping scenario.
MCRPL performs better than $\pi$-net and PSJ-net, indicating that our prompt module can effectively capture and transfer common preferences from dense domains. 
The gap between MCRPL and DAGCN fully demonstrates that MCRPL can leverage one or more dense domain information to generate more informative sequential representations in the target domain, thus addressing the issue of data sparsity. 
MCRPL performs better than PLCR, indicating that MCPRL can better adapt to sparse data scenarios.
6) We find that the performance of cross-domain methods is not as good as single-domain methods, which shows that under non-overlapping and data sparsity conditions, cross-domain methods that are based on overlapping entities may have negative transfer issues, hindering them from achieving better results.

\begin{table*}
  \caption{Ablation studies on the HVIDEO dataset, where E-domain is the sparse target domain. The best results on it are highlighted with bold text and the second-best result is highlighted with an underline.}
  \label{tab:Ablation}
  \centering
  \normalsize
  \resizebox{\textwidth}{!}{
    \begin{tabular}{cccrrrr}
      \toprule
      \multirow{2}{*}{Variants} & \multicolumn{6}{c}{\text{E-domain} $\leftarrow$ \text{V-domain} } \\
      \cmidrule{2-7}
      & Recall@3 & MRR@3 & Recall@5 & MRR@5 & Recall@10 & MRR@10 \\
      \midrule
      w/o domain orthogonal loss & \underline{17.023} & \underline{14.951} & \underline{20.062} & \underline{15.544} & \underline{26.469} & \underline{16.398} \\
      w/o Source Domain & 15.235 & 12.693 & 18.597 & 13.457 & 26.159 & 15.894 \\
      w/o Prompt & 16.846 & 14.072 & 19.231 & 14.782 & 25.180 &  16.032 \\
      w/o Two-stage Optimization & 16.291 & 13.862 & 19.632 & 14.617 & 26.559 & 15.550 \\
      w/o Prompt Attention & 15.972 & 13.951 & 19.540 & 14.272 & 26.471 & 16.358\\
      \textbf{\ac{MCRPL}} & \textbf{17.347} & \textbf{15.224} & \textbf{20.233} & \textbf{15.812} & \textbf{26.756} & \textbf{16.605}\\
      \bottomrule
    \end{tabular}
  }
\end{table*}

\begin{table*}
  \caption{Ablation studies on the MIXED dataset, where VIDEO is the spare target domain. The best results on it are highlighted with bold text and the second-best result is highlighted with an underline.}
  \label{tab:Ablation2}
  \centering
  \normalsize
  \resizebox{\textwidth}{!}{
    \begin{tabular}{ccccccc}
      \toprule
      \multirow{2}{*}{Variants} & \multicolumn{6}{c}{\text{VIDEO} $\leftarrow$ \text{(MOVIELENS, WIKIPEDIA)}} \\
      \cmidrule{2-7}
      & Recall@3 & MRR@3 & Recall@5 & MRR@5 & Recall@10 & MRR@10 \\
      \midrule
      w/o domain orthogonal loss & 4.123 & 3.091 & \underline{6.280} & 3.341 & \underline{9.842} & 3.895 \\
      w/o Source Domain & 3.231 & 2.181 & 4.860 & 2.595 & 8.556 & 3.373 \\
      w/o Prompt & 3.431 & 2.320 & 4.831 & 2.620 & 8.740 & 3.059 \\
      w/o Two-stage Optimization & 3.978 & 2.619 & 5.737 & 3.026 & 9.948 & 3.930 \\
      w/o Prompt Attention & \underline{4.125} & 2.947 & 5.893 & 3.085 & 9.657 & 3.871 \\
      w/o WIKIPEDIA & 3.894 & \underline{3.125} & 6.275 & \underline{3.425} & 9.473 & \underline{3.989} \\
      \textbf{\ac{MCRPL}} & \textbf{4.500} & \textbf{3.327} & \textbf{6.465} & \textbf{3.723} & \textbf{10.130} & \textbf{4.199} \\
      \bottomrule
    \end{tabular}
  }
\end{table*}

\section{Model Analysis}
In this section, we first conduct ablation studies to explore the importance of different components of \ac{MCRPL}, and then investigate the impact of different optimization strategies.
Finally, we report the impact of the model hyper-parameters and the training efficiency of \ac{MCRPL}.

\subsection{Ablation Study (\textbf{RQ2})}
To explore the importance of the components within \ac{MCRPL} and the impact of different training strategies, we compare \ac{MCRPL} with its following variants to answer RQ2:
\begin{itemize}
    \item w/o domain orthogonal loss: This is a variant of \ac{MCRPL} that removes the domain orthogonal loss $\mathcal{L}_{orthogonal}$ from $\mathcal{L}$. This is to demonstrate the importance of modeling different aspects of the domain information.
    \item w/o Source Domain: This is a variant of \ac{MCRPL} that directly trains \ac{MCRPL} on the target domain without any pre-training process.
    This is to evaluate the importance of pre-learned domain knowledge in the source domain.  
    \item w/o Prompt: This variant of \ac{MCRPL} removes the prompts within it, and
    only the parameters in the sequence encoder are pre-trained and fine-tuned.
    This is to explore the significance of the prompts in learning the common domain knowledge and adapting the pre-trained model to the target domain.
    \item w/o Two-stage Optimization: This is another variant of \ac{MCRPL} that only retains the pre-training stage of the training process. This is to demonstrate the contribution of the fine-tuning phase.
    \item w/o Prompt Attention: 
    This variant replaces the prompt attention module within \ac{MCRPL} an average pooling operation to demonstrate the importance of our attentive aggregation methods on prompt context words.
    \item w/o WIKIPEDIA (only for multiple source datasets): 
    This variant of \ac{MCRPL} explores the importance of the WIKIPEDIA dataset with MOVIELENS as its only source domain.
\end{itemize}

The comparison results with the above variants of \ac{MCRPL} are shown in Tables \ref{tab:Ablation} and \ref{tab:Ablation2}, from which we can observe that:
1) \ac{MCRPL} performs better than \ac{MCRPL} (w/o Prompt Attention) on both datasets, demonstrating the effectiveness of leveraging an attention mechanism in aggregating the prompt context words, which can fuse them by imposing different weights to the prompts.
2) \ac{MCRPL} achieves better results than \ac{MCRPL} (w/o Source Domain) on all datasets, indicating the importance of the pre-training phase. It can learn domains' common knowledge, and thereby significantly alleviates the data sparsity issue in the target domain. 
3) \ac{MCRPL} exhibits a significant improvement over \ac{MCRPL} (w/o Prompt), denoting 
the contributions of prompts in transferring the domain knowledge from the source to the target domain under the non-overlapping scenario.
With domain-agnostic and domain-related prompts, we can improve the recommendations in the target domain on the information in the source domains.
4) \ac{MCRPL} outperforms \ac{MCRPL} (w/o Two-stage Optimization) in all metrics, showing the importance of the fine-tuning stage, which serves as the adapter of adapting the pre-trained domain knowledge to the target domain.
5) \ac{MCRPL} performs better than \ac{MCRPL} (w/o Two-stage Optimization), indicating the essential of the domain orthogonal loss in learning the domain information from two different aspects, which enables the domain-agnostic and domain-related prompts to play different roles in our two-stage training process.
6) \ac{MCRPL} shows its advance over \ac{MCRPL} (w/o WIKIPEDIA) in the MIXED dataset, denoting the effectiveness of \ac{MCRPL} in modeling multiple source domains.
That is, \ac{MCRPL} has the ability to model multi-source domains, and can transfer the common knowledge to the target domain to improve the recommendations within it.

\begin{figure}
    \centering    \includegraphics[width=13cm]{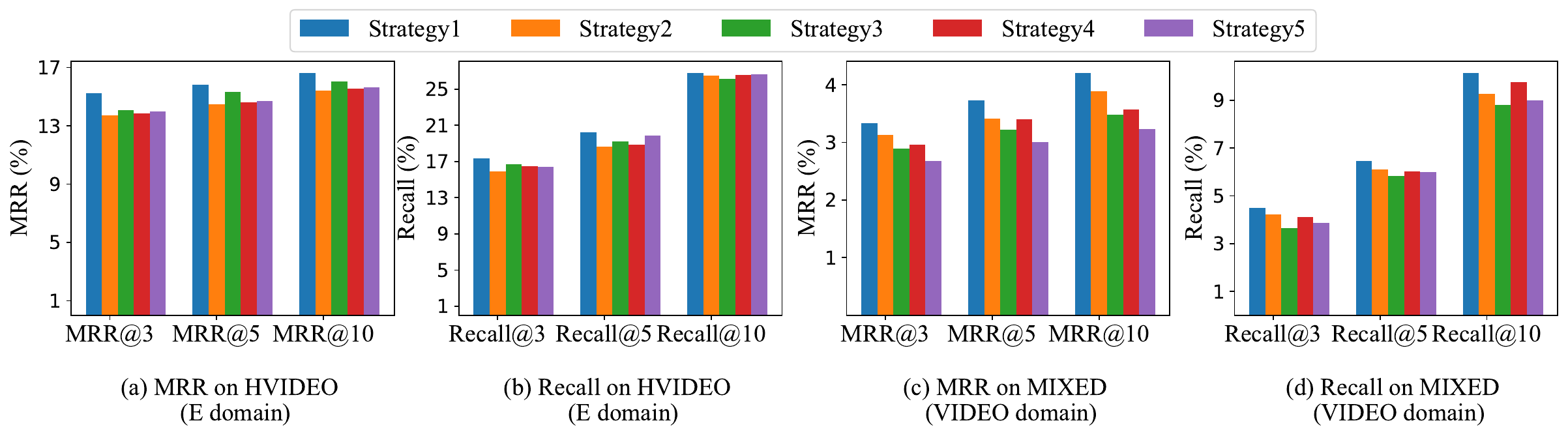}
    \caption{Impact of different fine-tuning strategies on both datasets.}
    \label{fig:optimization strategy}
\end{figure}

\subsection{Impact of Different Fine-tuning Strategies (\textbf{RQ3})}
To answer research question RQ3 and explore the impact of different fine-tuning strategies, we utilize $\boldsymbol{\theta}_{P_{c}}$, $\boldsymbol{\theta}_{P_{s}}$, $\boldsymbol{\theta}_{Emb}$, $\boldsymbol{\theta}_{Seq}$ and $\boldsymbol{\theta}_{Rec}$ to respectively represent the parameters in domain-agnostic prompts, domain-specific prompts, embedding layer, sequence encoder, and recommendation components.
Then, our fine-tuning strategies (from Strategy 1 to 5), and the corresponding parameter updates can be represented as follows:
\begin{itemize}
    \item Strategy 1: This method freezes the sequence encoder ($\boldsymbol{\theta}_{Seq}$), embedding layer ($\boldsymbol{\theta}_{Emb}$) and domain-agnostic prompts ($\boldsymbol{\theta}_{P_{c}}$), and optimizes the parameters in domain-specific prompts ($\boldsymbol{\theta}_{P_{s}}$) and recommendation components ($\boldsymbol{\theta}_{Rec}$).
    \item Strategy 2: In this method, we freeze the sequence encoder ($\boldsymbol{\theta}_{Seq}$) and  embedding layer ($\boldsymbol{\theta}_{Emb}$), and update domain-agnostic prompts ($\boldsymbol{\theta}_{P_{c}}$), domain-specific prompts ($\boldsymbol{\theta}_{P_{s}}$), and recommendation components ($\boldsymbol{\theta}_{Rec}$).
    \item  Strategy 3: This method freezes domain-agnostic prompts ($\boldsymbol{\theta}_{P_{c}}$), embedding layer ($\boldsymbol{\theta}_{Emb}$) and recommendation layer ($\boldsymbol{\theta}_{Rec}$), and optimizes the sequence encoder ($\boldsymbol{\theta}_{seq}$) and domain-specific prompts ($\boldsymbol{\theta}_{P_{s}}$).
    \item Strategy 4: This method updates the parameters in all components. 
    \item Strategy 5: This method switches between pre-train and fine-tune within every epoch. For its fine-tune strategy, we adopt the same parameters update and freezing methods as Strategy 1.
\end{itemize}

The comparison results are shown in Fig.~\ref{fig:optimization strategy}, from which we can have the following findings: 1) By freezing the parameters only in domain-agnostic prompts and sequence encoder (i.e., Strategy 1), we can achieve the best performance on all the datasets, indicating the importance of the common knowledge in guiding the recommender to adapt to the target domain.
2) Strategy 1 outperforms Strategy 2, demonstrating the essential of learning domain-specific prompts by fixing the domain-agnostic prompts. If we optimize them simultaneously, the usefulness of the prior knowledge induced in the pre-training stage will be greatly reduced.
3) Strategy 1 performs better than Strategies 3, and 4, indicating the harm of incorporating more parameters in the fine-tuning stage, which will result in the sub-optimal results in the target domain.
4) Strategy 1 performs better than Strategy 5, indicating the advantage of Strategy 1 in cross domain tasks, since the alternative optimization strategy may cause the interleaving of domain public information and domain private information.

\begin{figure}
    \centering    \includegraphics[width=10cm]{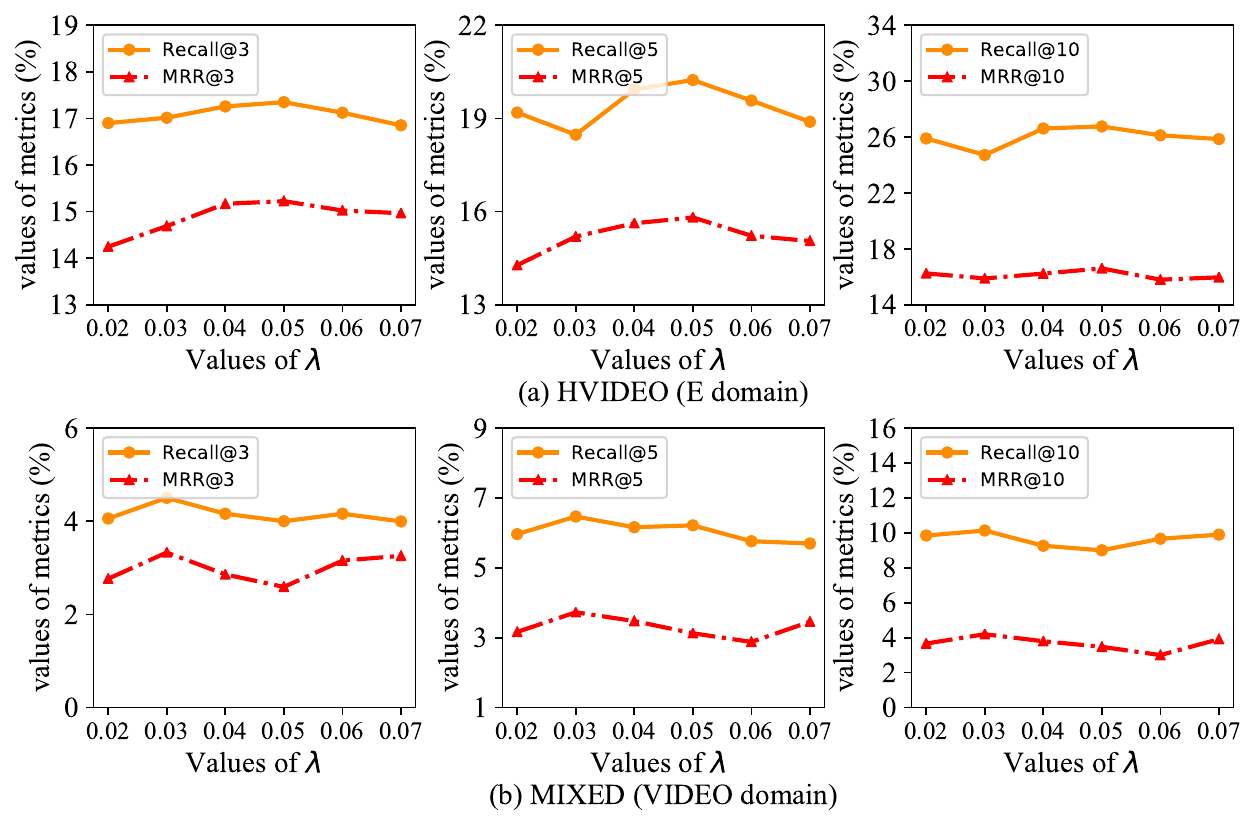}
    \caption{Impact of the hyper-parameter $\lambda$ on HVIDEO and MIXED datasets.}
    \label{fig:WSDM-hyper}
\end{figure}

\begin{figure}[h]
    \centering
    \includegraphics[width=10cm]{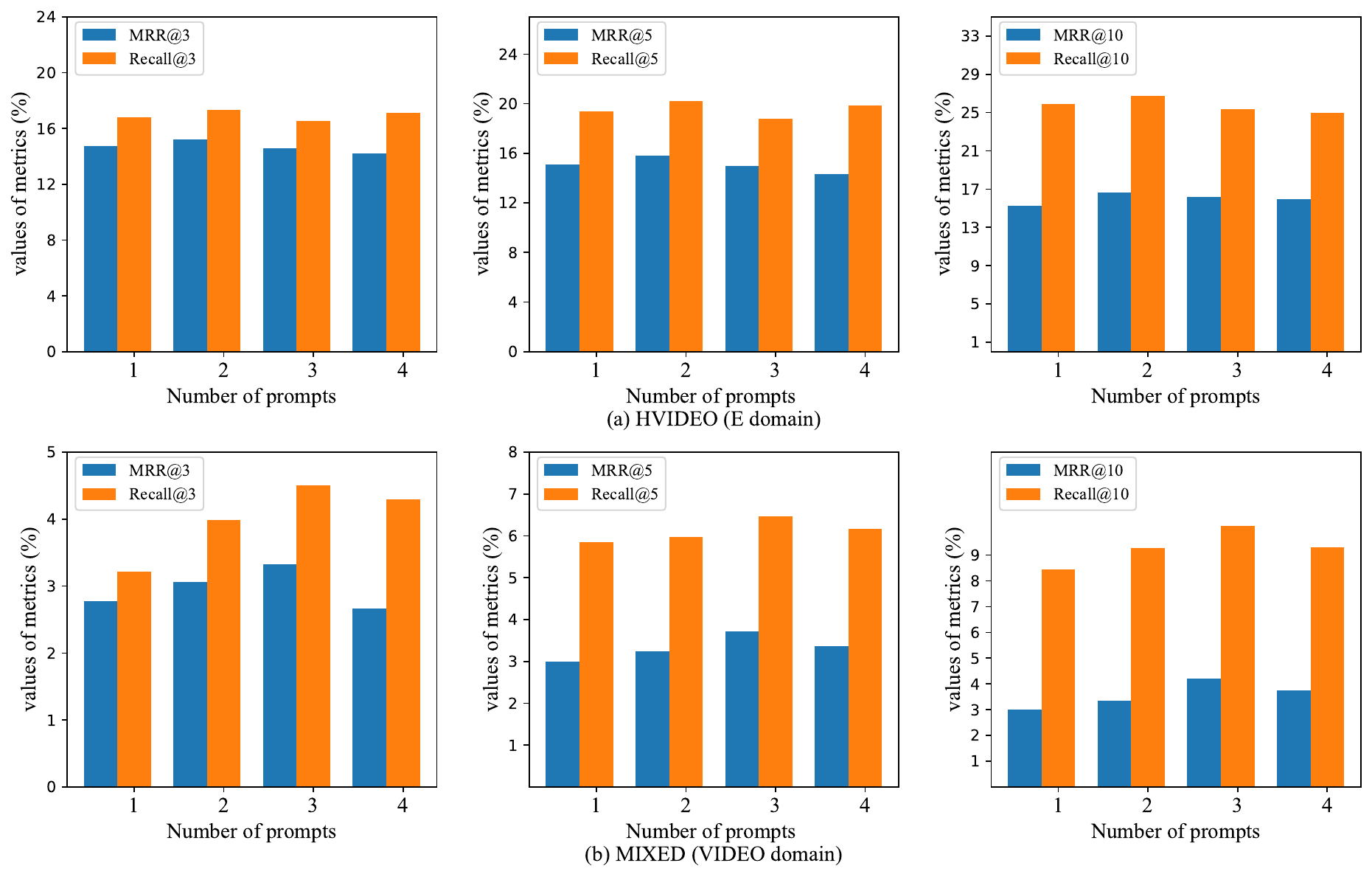}
    \caption{ Impact of the prompt length $L_{p}$ on HVIDEO and MIXED datasets.}
    \label{fig:number}
\end{figure}

\subsection{Hyper-parameter Analysis (\textbf{RQ4})}

To show the impact of the key hyper-parameters on the performance of \ac{MCRPL}, we further conduct experiments on all datasets by analyzing $\lambda$ and $L_p$, and report their results in Figs.~\ref{fig:WSDM-hyper} and \ref{fig:number}.

\begin{figure}[h]
    \centering    \includegraphics[width=13cm]{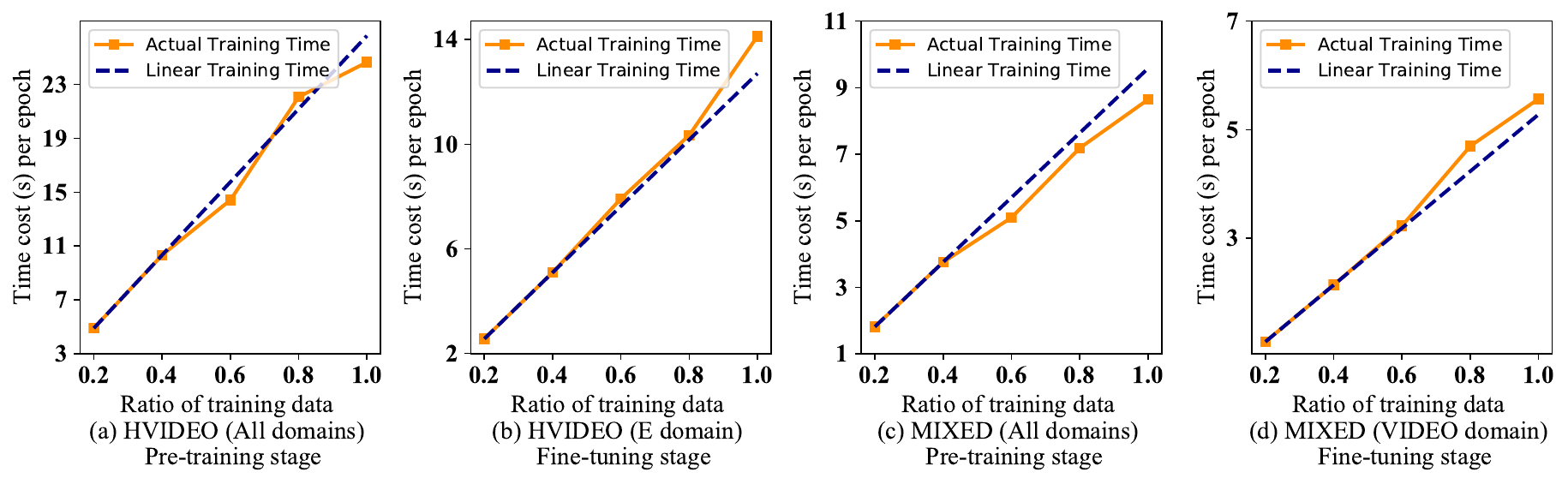}
    \caption{ Training efficiency and scalability of \ac{MCRPL} on HVIDEO and MIXED.}
    \label{fig:Training Efficiency}
    \vspace{-3mm}
\end{figure}

The hyper-parameter 
$\lambda$ controls the importance of the domain orthogonal loss in fine-tuning \ac{MCRPL}. 
From the experimental results shown in 
Fig.~\ref{fig:WSDM-hyper} we can find that when $\lambda$ increases from 0.02 to 0.07, the performance of \ac{MCRPL} fluctuates within a small range, indicating the insensitivity of $\lambda$ to \ac{MCRPL}. And, we can get better results with a probably $\lambda$ value, that is, we can model the domain information in different aspects.
The hyper-parameter $L_p$
denotes the number of context words in prompts. The experimental results are shown in Fig.~\ref{fig:number}, from which we have the following observations:
1) We can achieve the best performance of \ac{MCRPL} when set $L_p$ as 2, and the model performance gradually declines with $L_p$ exceeding 2. This is because HVIDEO has a relatively small number of items, and when the prompt length is 2, it is already sufficient in capturing the common information in the dense domain. Having too many prompts can introduce noise and result in sub-optimal performance.
2) \ac{MCRPL} performs the best when the $L_p$ is set as 3 on the MIXED dataset. We speculate that this is because the MIXED dataset has a larger number of items than HVIDEO. It needs a larger value of $L_p$ to adequately capture the common information from different aspects in the dense source domain.

\subsection{Training Efficiency and Scalability (RQ5)}
We evaluate the training efficiency and scalability of \ac{MCRPL} by measuring its training cost in different data ratios on both datasets. We split the training dataset into $\{0.2, 0.4, 0.6, 0.8, 1.0\}$ while keeping all the hyper-parameters fixed. Fig.~\ref{fig:Training Efficiency} shows the training cost of \ac{MCRPL} compared with the linear training time. From this result, we can find that the cost of our training process is linearly associated with the data amount. The training cost almost linearly grows (from $0.005\times 10^3$ to $0.024\times10^3$ and $0.003\times 10^3$ to $0.014\times10^3$  in both the pre-training and fine-tuning stages on HVIDEO, from $0.002 \times 10^3$ to $0.009\times10^3$ and $0.001 \times 10^3$ to $0.006\times10^3$ in both the pre-training and fine-tuning
stages on MIXED) with the increment of the training data on all datasets, which demonstrates the training scalability of \ac{MCRPL} in large-scale datasets.
\section{Conclusions}

In this work, we focus on the \ac{NMCR} task and propose \ac{MCRPL} to address it, where a prompt-enhanced sequence encoder along with a two-stage training schema is developed. 
Specifically, to learn transferable domain information without any overlapping information, we encode the domains' common knowledge by devising the domain-agnostic and domain-specific prompts in the pre-training stage, which are taken as the enhancement of items with a prompt attention mechanism.
Then, to transfer the domain knowledge from the source to the target domain, we leverage the fine-tuning strategy by updating the domain-specific prompts, while keeping the domain-agnostic prompts and sequence encoder fixed to preserve domain-specific information in the target domain with the guidance of the pre-learned domain knowledge.
To enable the domain-agnostic and domain-specific prompts can model different aspects of the domain information, an orthogonal constraint is added to the target loss.
We conduct extensive experiments on the HVIDEO and MIXED datasets, 
and the results show the advance of \ac{MCRPL} over related SOTA methods.
Our MCRPL method can also be applied to rating prediction and click-through rate prediction tasks.
The key point our method is to embed user preference into common prompts, which can be shared across domains, thus to improve recommendations in the target domain. This target is irrelevant to the task itself.

One limitation of MCRPL is that it focuses on modeling users by only exploiting their historical interactions, while the auxiliary information, such as comments, reviews, and description information is not investigated, which may also lead us to discover powerfully cross-domain recommendation algorithms.
We will take this study as one of our future works.


\begin{acks}
This work was supported by the National Natural Science Foundation of China (Nos. 62372277, 62172263, 62002209), Natural Science Foundation of Shandong Province (Nos. ZR2022MF257, ZR2020YQ47, ZR2022MF295, ZR201911130645, ZR2020QF111), CCF-Baidu Open Fund (No. CCF-BAIDU OF2022008), Australian Research Council Future Fellowship (No.FT210100624), and Humanities and Social Sciences Fund of the Ministry of Education (No. 21YJC630157).
\end{acks}

\bibliographystyle{ACM-Reference-Format}
\bibliography{sample-base}










\end{document}